# ROBUST VARIABLE SELECTION FOR HIGH-DIMENSIONAL REGRESSION WITH MISSING DATA AND MEASUREMENT ERRORS

**Zhenhao Zhang**
China University of Petroleum
College of Computer Science and Technology
zhangzhh2002@gmail .com

**Yunquan Song***
China University of Petroleum
College of Science
statistics99@163 .com

## ABSTRACT

In our paper, we focus on robust variable selection for missing data and measurement error.Missing data and measurement errors can lead to confusing data distribution.We propose an exponential loss function with tuning parameter to apply to Missing and measurement errors data.By adjusting the parameter,the loss function can be better and more robust under various different data distributions.We use inverse probability weighting and additivity error models to address missing data and measurement errors. Also, we find that the Atan punishment method works better.We used Monte Carlo simulations to assess the validity of robust variableselection and validated our findings with the breast cancer dataset

## 1 Introduction

The linear relationship between response variables and covariates has been the topic of interest. In the classical squared loss function, it is usually assumed that the data obey a normal distribution.However, the data discussed in this paper contain a large number of missing data and measurement errors, such that the data usually do not conform to any of the common forms of data distribution. We propose a method based on an exponential squared loss function with tuning parameter. For data with different distributions, a better result of linear regression can be achieved by changing the value of the tuning parameter $h$. Therefore, for any kind of data distribution, going with an exponential squared loss function with moderating variables will be highly robust.For any data distribution, the loss function is strongly robust for $h \in (0, +\infty)$.In previous studies, when using the traditional squared loss function, the data distribution requirements are very high, resulting in the traditional exponential squared loss function being very sensitive to anomalies. This reduces the estimation efficiency of the model, and this drawback becomes more obvious in data containing missing data with measurement errors.In contrast, the use of exponential squared loss functions can improve the estimation efficiency of the model by varying the tuning parameter $h$in a way that adapts to more distributed forms of data sets and produces more reliable estimates.

In the traditional squared loss function, the values of the covariates are always defaulted to be free of missing data and measurement errors. Even if missing data and measurement errors exist, they are assumed to be absent or these data are removed.However, this assumption is often broken in studies in disciplines such as health and epidemiology.As an illustration, Zhang and Zhou(1) looked at a collection of breast cancer patients to identify the gene expression that was associated with long-term disease-free survival. The data collection consists of 24481 gene probes collected from 78 breast cancer patients. In particular, using the log-value of the ratio $(\log_{10}(\text{Ratio}))$, which could be denoted as $Y$, it is possible to forecast the disease-free survival.In truth, gene sensors will inevitably lead to measurement errors.In this breast cancer data set, the $(\log_{10}(\text{Ratio}))$ numbers have missing data.

When there are a large number of missing data and measurement errors in a dataset, if we ignore the missing data and measurement errors and use the traditional square loss function for estimation, the estimation accuracy of the model will be greatly affected due to the chaotic data distribution, resulting insignificant estimation bias. In the above dataset, We discover that employing the traditional squared loss function, which handles data with measurement errors and missing data, is subpar and decreases the validity of the estimate process. With the help of the tuning parameter $h$, the exponential squared loss can be used to solve this issue.

In observational studies, the statistical method of inverse probability weighting (IPW) is often used to estimate causal effects. Because treatment assignment in observational studies is not randomized, there is a potential for selection bias.The IPW approach addresses this problem by weighting the data to reflect the likelihood of receiving treatment or a control group. The use of the IPW method allows estimation of the average treatment effect (ATE) as if the treatment assignment were randomized. In reality, data are often missing for a variety of unpredictable reasons. Assuming that these missingnesses do not exist when building regression models can hinder the validity of the study and lead to important errors. In this paper, we use the IPW approach to address the problem of missing data.

For the measurement error, we use a simple and effective additive error model. In the classical additive measurement error model, Rosner et al.(2). proposed a simple regression correction solution to correct for measurement error because the technique does not make assumptions about the parameter error distribution and therefore cannot be used directly for linear regression.He and Liang(5) considered a class of orthogonal regression methods with linear quantile models of measurement error.

The rest of the paper is organized as follows. Section 2 gives the linear regression model with an exponential squared loss function with conditioning parameters. Section 3 describes methods to deal with missing data and measurement errors. The various penalties used to perform variable selection on the model are described in Section 4. Experiments with simulated and real data are conducted in Section 5 to confirm our conclusions.

## 2 Linear regression with exponential squared loss

The issue we explore in this paper centers on a linear regression model. We presum that the variables in this model are the response variable $Y_i$ and the covariate $X_i = (X_{i1}, X_{i2}, \dots X_{id})$, where $X_i$ is a $d$-dimensional vector and Y and X have a linear connection. The model we're interested in can be expressed as:

$$Y_i = X_i^T \omega + b_i, \quad i = 1, 2, \dots n, \tag{I}$$

where $b_i$ is the errors term and $\omega = (\omega_1, \omega_2, \dots, \omega_d)$ is the set of unknown values to be estimated. near connection between the covariates and the response variable.Our goal is to use the sample to determine the unknown parameters.This research focuses on the non-traditional loss function we suggest for high-dimensional datasets with missing values and measurement errors. This loss function can be expressed as:

$$\min_{\omega} \quad \sum_{i=1}^{n} 1 - exp(-(Y_i - {X_i}^T \omega)^2/h)$$

Based on the tuning parameter $h$, this loss function is unique from others.By altering the value of the tuning parameter $h$, the loss function can be made to work better for different data distributions.We will discuss the influence on the loss function when $h$ has different values below.

When the value of the tuning parameter $h$is large, We consider $-(Y_i - X_i^T\omega)^2/h$ to be infinitesimal.There are the following transformation formulas based on knowledge of the limit.

$$\exp(-(Y_i - X_i^T\omega)^2/h) \approx -(Y_i - X_i^T\omega)^2/h + 1$$

According to the above formula,We can infer the following conclusion

$$1 - \exp(-(Y_i - X_i^T\omega)^2/h) \approx (Y_i - X_i^T\omega)^2/h$$

The loss function in this case is similar to that of the least squares function, which is a highly classic loss function with good results in many classic data distribution situations.

When the value of the tuning parameter $h$is small,The absolute value of $(Y_i - X_i^T\omega)$/h will become very large,and $\exp(-(Y_i - X_i^T\omega)^2/h)$ will be very small.Missing data and measurement errors often lead to the appearance of anomalous data. When there are outliers or outliers in the data, these anomalies do not cause the loss function to be less effective, reflecting the strong robustness of the loss function

From the above example, it is clear that the loss function has strong robustness since we can alter the value of the tuning parameter $h$ to make the loss function have a better impact whether the data distribution resembles the traditional data distribution or not. In reality, the loss function may get the optimum results for many data distributions by choosing an ideal tuning parameter $h$ value, which is also a reflection of the great robustness of loss function

In the situation covered in this work, measurement mistakes and missing data might cause the distribution of data to typically deviate from the standard distribution form.When utilizing the conventional loss function, the outcomes will be subpar.When the loss function is used with the tuning parameter $h$, the value of the tuning parameter $h$ can be changed to improve the performance of the loss function and increase the robustness of the function.

## 3 Missing data and measurement errors in linear regression

When model (I) contains missing data.We assume that $X_i^{(p)} \in R^m$ contains no missing data and $X_i^{(q)} \in R^n$ contains missing data.We define an indicator variable $F_i$. When $X_i^{(q)}$ does not contain missing data, the indicator variable $F_i$is 1. When $X_i^{(q)}$ contains missing data, the indicator variable $F_i$ is 0.

Missing data is any information that is not available or not collected in certain situations in a data set. Missing data can occur for a variety of reasons, such as missing data measurements, loss of data during storage or transmission, etc. Missing data can be classified into different types based on the underlying mechanism causing the missingness. For example, missing completely at random (MCAR) occurs when the probability of missingness is unrelated to both observed and unobserved data. Missing at random (MAR) occurs when the probability of missingness is related to observed data but not to unobserved data. Finally, missing not at random (MNAR) occurs when the probability of missingness is related to unobserved data.

Since the measurement and transmission of various data are random and we need to model the missing data.In this paper, we consider that the data are missing at random (MAR).The probability of a vector of variables containing missing data is dependent on the variables that never contain missing data rather than those that may contain missing data. Consequently, the following corollary can be obtained

$$\begin{aligned} \pi_{i0} &\equiv Pr(F_i = 1|Y_i, X_i) \\ &= Pr(F_i = 1|Y_i, X_i^{(p)}) \\ &= Pr(F_i = 1|s_i) \end{aligned}$$

For the missing data,We consider the Inverse probability weights(IPW).IPW is a method assigns different weights to the observations that never contain missing data, in order to alleviate potential bias. The probability $\pi_{i0}$ is often calculated using logistic or probit regressions. The smooth kernel is taken into account by Chen et al.(3).In their study using nonparametric kernel averaging.

$$\tilde{\pi}_{i0} = \frac{\sum_{j=1}^{n} F_j M_l(s_i - s_j)}{\sum_{j=1}^{n} M_l(s_i - s_j)}$$

where $l$ stands for the frequency variable and $M_h(\cdot) = M(\cdot/l)/l^{m+1}$designates a kernel function with $(m+1)$ variables. The typical techniques for choosing a bandwidth. are SROT,NROT, BCV, and LSCV, as well as automated bandwidth selection. We use the streamlined Sepanski et al(4) approach to choose bandwidth in the experiment and take into account Gaussian kernels.

As we conduct the experiment, we consider the Gaussian kernel and select the bandwidth by utilizing a condensed form of Sepanski et al(4).The bandwidth $l$is determined as $\sigma n^{-1/(m+1+2)}$,where $\sigma$ is the sample standard deviation of covariates $(Y_i, T^{(p)})^T$ in the model, $T^{(p)}$ is a stand-in for $X^{(p)}$, and 2 is taken from the order of the Gaussian kernel.To simplify the notation, we designate $\pi_i 0$ as the actual chance that observation i contains all of the data, and $\tilde{\pi}_i$ as the anticipated probability using the kernel notation as a foundation. Consequently, the following is a description of the regression estimator:

$$\hat{\omega}_X = \arg\min_{\omega_X} \sum_{i=1}^{n} \frac{F_i}{\tilde{\pi}_i} \left(1 - exp(-(Y_i - X_{i,X}{}^T \omega_X)^2/h)\right)$$

A traditional additive measurement errors model is taken into consideration when $X_i$ is measurement errors.

$$T_i = X_i + G_i$$

It is frequently presumed that the measurement errors $G_i$ and the model errors $b_i$ in the measurement errors model have normal distributions, and that the alternative $T_i$ was observed.The variables $X_i$ and $G_i$ are further assumed to be independent of one another.

To deal with specific linear models that have measurement mistakes,He and Liang(5) suggested the orthogonal regression technique.In this case,To manage the linear model with measurement mistakes and missing data,orthogonal regression and IPW will be used.Let $T_{i,X}$ and $G_{i,X}$ stand for the measured values and measurement errors connected to the factors $X_{i,X}$,respectively. The ability of orthogonal regression to predict the coefficients in a linear model has been demonstrated. The loss function can consider as:

$$\tilde{\omega}_X = \arg\min_{\omega_X} \sum_{i=1}^{n} \frac{F_i}{\tilde{\pi}_i} \left( \frac{1 - exp(-(Y_i - T_{i,X}{}^T \omega_X)^2/h)}{\sqrt{1 + \|\omega_X\|^2}} \right) \tag{2}$$

where $\frac{Y_i - T_{i,X}^T \omega_X}{\sqrt{1+\|\omega_X\|^2}}$ is the orthogonal residual.In linear models with additive errors, the adjustment factor $\sqrt{1 + \|\omega_X\|^2}$ is frequently used, and $\|\cdot\|$ stands for the vector's $L_2$ norm.

## 4 Nonconvex penalized estimation

The problems discussed in this paper address high-dimensional data,due to the singularity of the information matrix, the loss function minimization method is not very simple.Since it is not known which variables are important in high-dimensional data, in order to make the coefficient matrix sparse and easy to solve, we can select the variables by adding a penalty term to the loss function.

For punishment method, the main commonly used punishment method are Adaptive Lasso, SCAD, $l_p$ and MCP are a few examples where $l_p$ is widely used.Typically, the number of p is an integer between 0 and 2.However,because the loss function is irregular at 0,it is unstable to use $l_0$ norm as the penalty function,and for modelselection,$l_1$ might be unreliable and prejudiced. For the adaptive Lasso to achieve the oracle property, it modifies its weight.The smoothly clipped absolute deviation (SCAD) penalty,a continuously differentiable non-convex punishment component, was explored by Fan and Licitep(6).Zhang (7) suggested the minimax concave penalty (MCP), a different non-convex penalty function that has the ability to choose the right model with a probability that is close to 1.

In a related work, The Atan non-convex penalty function was introduced by Wang and Zhu(8), and demonstrated how it generates an Atan estimator with a number of beneficial theory characteristics, such as unbiasedness and sparsity.These additions broaden the selection of penalty functions that can be used in statistical modeling and present new research directions.

The Atan punishment is described as follows:

$$p_f(|x|) = f\left(u + \frac{2}{\pi}\right) \arctan\left(\frac{|x|}{u}\right)$$

for $f \geq 0$ and $u > 0$.The following loss function, which includes a penalty term that reduces some coefficient components to zero, is suggested as a way to account for missing data and measurement errors.

$$\hat{\omega}_f = \arg\min_{\omega} \sum_{i=1}^{n} \frac{F_i}{\tilde{\pi}_i} \left( \frac{1 - exp(-(Y_i - T_{i,X}{}^T \omega_f)^2/h)}{\sqrt{1 + \|\omega_f\|^2}} \right) + \sum_{k=1}^{p_n} p_f\left(|\omega_f|\right) \tag{3}$$

where $\omega_f$is the coefficient to be estimated,$F_i$means whether it contains missing values,$\pi_i$represents the probability of occurrence,$Y_i$is the response variable,$T_{i,x}$is the covariate with error,$h$is the tuning parameter,$p_f(\cdot)$ be the Atan punishment method

$$p_f(|\omega_k|) = f\left(u + \frac{2}{\pi}\right) \arctan\left(\frac{|\omega_k|}{u}\right)$$

and the nonnegative punishment method f.

The optimization problem of formula (3) is a minimization problem, and a common solution to this problem is numerical iteration. In this paper, we use the most common gradient descent method to solve this problem

In the model, the selection of punishment method $f$ is particularly important.A BIC method for high-dimensional variables is suggested by Lee et al(9).

Using the punishment method $f$, let $\hat{\omega}_f = (\hat{\omega}_{f,1}, \ldots, \hat{\omega}_{f,d})$ be the penalized estimator. Let $S_f = \{1 \leq k \leq d : \hat{\omega}_{f,k} \neq 0\}$ be the Atan estimator's index collection of nonzero coefficients with punishment method $f$ .We obtain the subsequent solution.

$$\mathrm{HBIC}(f) = \log\left(\sum_{i=1}^{n} \frac{F_i}{\tilde{\pi}_i}\left(\frac{1 - exp(-(Y_i - T_{i,X}{}^T\omega_f)^2/h)}{\sqrt{1 + \|\hat{\omega}_f\|^2}}\right)\right) + |S_f|\,\frac{\log(\log(n))}{n}E_n$$

where $|S_f|$ represents the cardinality of $S_f$, and $E_n$ is a series of positive numbers infinite as $n$ increment.We calculate the value off by minimizing the objective function $HBIC(f)$ .

# 5 Experiment

## 5.1 Generate data example

In this experiment, we use Monte Carlo simulation to compare the effects of several punishment methods.We use four different punishment items, namely Lasso, Atan, MCP and SCAD.We generate 300 data sets, each data set contains 100 data, and each data has 300 or 500 dimensions.We generate covariates with normal distribution, with mean value of 0 and variance of 1, and randomly generate an errors term.To investigate how covariate association affects the choice of variables, we set up simulation experiments from the following two aspects:(1)Any two covariates are independent of each other,$r(X_i, X_j) = 0$ (2)Covariate relationships progressively decreases,$r(X_i, X_j) = 0.5^{|i-j|}$ .For coefficient matrix $\omega$,Let $\omega = (0, 1, 0, 2, 0, 4, \underbrace{0, \ldots, 0}_{d-6})$.We use the following methods to generate data

$$Y_i = X_{i2} + 2X_{i4} + 4X_{i6} + \varepsilon_i$$

We let the errors also conform to certain distribution conditions, mainly considering the following three cases:(1)regular distribution in general(2)three-degrees-of-freedom $t$ distribution (3)two-degrees of freedom $v$ distribution.

In our experiment,We assume that $X_1$, $X_3$, and $X_5$ possibly contain missing data during the experiment. The model is taken into account:

$$P(Pr(F_i) = 1) = 1 + 2Y_{i2} - 2X_{i3} + 4X_{i5}$$

We take into consideration the traditional additive model, $T_i = X_i + G_i$, for the measurement errors model.Where $G_i$ has a mean value of 0 and a variation of 0.3, satisfying the normal distribution.

In the following four conditions, we investigate the effect of various punishment method.
(1)Taking into account both measurement errors and missing data
(2)Only measurement errors is considered
(3)Only consider data missing
(4)Measurement errors and data missing are not considered

We use Ind to indicate that there is no association between the covariates. Instead, we use corr to indicate that there is an association between covariates.In Table 1 and Table 2, we calculate the model errors under different penalty punishment method and different covariate correlations.The error can be expressed as $\|\hat{\omega} - \omega\|^2$.When we ignore measurement errors and missing data, this because of the incomplete data with large errors, the value of the loss function becomes large.In the same case, when the error distribution satisfies $N(0, 1)$ and $t(3)$, the value of the loss function is smaller than chisq(2) because chisq(2) is an asymmetric distribution, which violates our assumptions.In summary, it is essential to deal with missing data and measurement errors.

The data in the Table 1 and Table 2 can be used to make the following conclusions: The results obtained using the model suggested in this paper are frequently better than when only measurement errors are considered, only data missing is considered, and measurement errors and data missing are not considered.

The effects of various punishment method can be obtained from Table 1 and Table 2.We found the following conclusions:(1)The exponential square loss function containing the tuning parameter $h$ has strong robustness.(2)Atan punishment method is more effective in selecting suitable variables; Lasso punishment method and other punishment method have selected effective variables.(3)The error value of the Atan penalty term will be lower compared to other penalty forms.(4)Regardless of the loss function and penalty term used, the number of selected covariates is still higher than the actual number of valid covariates.

Table 1: n=100,d=300

| 2*Error | 2*Covarites | lasso h=0.1 | 1 | 10 | scad h=0.1 | 1 | 10 | mcp h=0.1 | 1 | 10 | atan h=0.1 | 1 | 10 |
|---|---|---|---|---|---|---|---|---|---|---|---|---|---|
| Taking into account both measurement errors and missing data | | | | | | | | | | | | | |
| N(0,1) | Corr | 4.429 | 9.745 | 10.982 | 4.511 | 4.536 | 4.558 | 4.548 | 5.964 | 6.616 | 4.562 | 6.107 | 6.226 |
| | Ind | 4.445 | 9.477 | 10.792 | 4.504 | 4.531 | 4.554 | 4.551 | 5.963 | 6.557 | 4.560 | 6.199 | 6.585 |
| t(3) | Corr | 4.499 | 10.751 | 10.973 | 4.537 | 4.554 | 4.569 | 4.577 | 6.366 | 6.674 | 4.837 | 6.383 | 6.570 |
| | Ind | 4.476 | 10.529 | 10.951 | 4.528 | 4.546 | 4.564 | 4.567 | 6.265 | 6.622 | 4.577 | 6.280 | 6.638 |
| chisq(2) | Corr | 4.428 | 10.000 | 10.931 | 4.511 | 4.533 | 4.556 | 4.549 | 6.140 | 6.619 | 4.563 | 6.136 | 6.587 |
| | Ind | 4.432 | 9.360 | 10.832 | 4.504 | 4.531 | 4.555 | 4.554 | 6.023 | 6.629 | 4.565 | 6.226 | 6.586 |
| Only measurement errors is considered | | | | | | | | | | | | | |
| N(0,1) | Corr | 5.234 | 10.322 | 11.165 | 4.751 | 5.423 | 4.587 | 5.038 | 6.132 | 7.595 | 5.275 | 6.607 | 6.697 |
| | Ind | 5.204 | 10.122 | 10.901 | 4.543 | 5.094 | 4.273 | 4.892 | 5.916 | 7.587 | 4.783 | 6.524 | 6.644 |
| t(3) | Corr | 4.559 | 11.433 | 11.015 | 4.608 | 5.076 | 4.666 | 5.395 | 7.184 | 7.396 | 4.987 | 7.043 | 7.089 |
| | Ind | 4.372 | 11.334 | 10.771 | 4.439 | 4.600 | 4.206 | 5.369 | 6.815 | 7.262 | 4.775 | 6.769 | 6.617 |
| chisq(2) | Corr | 5.401 | 10.649 | 11.731 | 4.965 | 4.965 | 5.381 | 4.632 | 6.273 | 6.792 | 4.954 | 6.967 | 7.390 |
| | Ind | 5.192 | 10.157 | 11.581 | 4.614 | 4.632 | 5.112 | 4.283 | 5.940 | 6.703 | 4.890 | 6.468 | 7.305 |
| Only missing data is considered | | | | | | | | | | | | | |
| N(0,1) | Corr | 5.267 | 10.883 | 12.047 | 5.420 | 5.613 | 4.956 | 5.499 | 7.114 | 7.751 | 6.131 | 7.252 | 7.073 |
| | Ind | 5.054 | 10.727 | 11.966 | 5.331 | 5.402 | 4.909 | 5.199 | 6.878 | 7.403 | 5.781 | 6.932 | 7.056 |
| t(3) | Corr | 4.750 | 11.861 | 11.497 | 4.729 | 5.666 | 4.892 | 5.780 | 7.767 | 7.648 | 5.277 | 7.660 | 7.354 |
| | Ind | 4.716 | 11.701 | 11.232 | 4.401 | 5.462 | 4.482 | 5.420 | 7.283 | 7.382 | 5.115 | 7.607 | 7.049 |
| chisq(2) | Corr | 6.225 | 11.632 | 12.461 | 5.309 | 5.549 | 5.489 | 5.538 | 7.153 | 7.610 | 5.215 | 7.561 | 7.413 |
| | Ind | 5.836 | 11.420 | 12.416 | 5.176 | 5.472 | 5.348 | 5.318 | 6.889 | 7.381 | 4.777 | 7.302 | 6.941 |
| Measurement errors and missing data are considered | | | | | | | | | | | | | |
| N(0,1) | Corr | 8.354 | 16.622 | 17.590 | 8.135 | 8.494 | 7.661 | 8.424 | 9.985 | 11.701 | 8.723 | 10.895 | 10.961 |
| | Ind | 8.263 | 16.278 | 17.489 | 7.848 | 8.423 | 7.553 | 8.390 | 9.769 | 11.444 | 8.518 | 10.735 | 10.719 |
| t(3) | Corr | 7.240 | 18.056 | 17.390 | 7.008 | 8.508 | 7.459 | 9.068 | 11.214 | 11.630 | 8.016 | 11.373 | 11.410 |
| | Ind | 6.997 | 17.801 | 17.152 | 6.653 | 8.430 | 7.193 | 8.980 | 11.169 | 11.402 | 7.847 | 11.201 | 11.162 |
| chisq(2) | Corr | 8.961 | 17.299 | 18.498 | 7.732 | 8.017 | 8.694 | 7.983 | 10.184 | 11.057 | 8.082 | 11.040 | 11.656 |
| | Ind | 8.672 | 17.168 | 18.249 | 7.576 | 7.702 | 8.381 | 7.886 | 9.954 | 10.839 | 7.880 | 10.714 | 11.557 |

### 5.2 Real data example

In this section, we use some real data for our experiments. We illustrate the proposed method by analyzing classical breast cancer data. This dataset has been widely used in previous studies, which means that it is highly representative. This dataset can be found at https://ccb.nki.nl/data/van-t-Veer_Nature_2002/. In 98 breast cancer tissues, 24,481 different gene expressions have been identified.

We observed that the Ratio gene showed a deletion of expression. During actual medical measurements, measurement errors can occur frequently, which leads to a high number of measurement errors in our data. In addition, according to biological knowledge, not every gene affects the expression of a trait. Therefore, we will use our model to select out the dominant genes that affect the expression of the trait. Therefore, we used disease-free survival time as the response variable $Y_i$ and the the log-value of the ratio ($\log_{10}(\text{Ratio})$), as the covariate $X_i$.

Based on the existing biological knowledge, our hypothesis can be easily disproved.The number of genes associated with breast cancer is only a small fraction of all genes.Therefore, our task remains to perform variableselection on high-dimensional data.We modify the tuning parameter h=(0.1, 1, 10) to equal various punishment method and add them to the loss function.

For each different parameter h, we randomly selected 3000 genes for the simulation experiment, and then calculated the correlation coefficient between these 3000 genes and disease-free survival time, and selected the top five genes with the largest correlation coefficient.We put the comparison results in Table 3.

Similarly, when conducting experiments on real data, we also conducted experiments on the following three situations: (1) Only measurement errors are considered (2) Only consistent data missing (3) Measurement errors and data missing are not considered.

Table 4 shows that our proposed method has less error and better robustness of the model. Thus, our method is an effective and robust method in dealing with data with missing data and high-dimensional covariate measurement errors.

Table 2: n=100,d=500

| **2*Error** | **2*Covarites** | **lasso h=0.1** | **1** | **10** | **scad h=0.1** | **1** | **10** | **mcp h=0.1** | **1** | **10** | **atan h=0.1** | **1** | **10** |
|---|---|---|---|---|---|---|---|---|---|---|---|---|---|
| Taking into account both measurement errors and missing data | | | | | | | | | | | | | |
| N(0,1) | Corr | 4.567 | 10.453 | 10.837 | 4.566 | 6.939 | 7.023 | 4.567 | 10.453 | 10.837 | 4.567 | 10.453 | 10.837 |
| | Ind | 4.565 | 10.318 | 10.982 | 4.565 | 6.726 | 7.010 | 4.565 | 10.318 | 10.982 | 4.565 | 10.318 | 10.982 |
| t(3) | Corr | 4.576 | 10.846 | 10.975 | 4.578 | 79.128 | 287.620 | 4.576 | 10.846 | 10.975 | 4.576 | 10.846 | 10.975 |
| | Ind | 4.578 | 10.747 | 10.987 | 4.578 | 77.657 | 494.196 | 4.578 | 10.747 | 10.987 | 4.578 | 10.747 | 10.987 |
| chisq(2) | Corr | 4.561 | 10.352 | 10.888 | 4.561 | 6.813 | 6.947 | 4.561 | 10.352 | 10.888 | 4.561 | 10.352 | 10.888 |
| | Ind | 4.562 | 10.359 | 10.914 | 4.562 | 6.817 | 6.984 | 4.562 | 10.359 | 10.914 | 4.562 | 10.359 | 10.914 |
| Only measurement errors is considered | | | | | | | | | | | | | |
| N(0,1) | Corr | 5.628 | 11.294 | 11.799 | 5.235 | 7.405 | 8.383 | 5.272 | 11.868 | 12.336 | 5.404 | 11.632 | 12.221 |
| | Ind | 5.449 | 11.160 | 10.690 | 4.897 | 7.050 | 7.847 | 4.985 | 11.100 | 11.665 | 5.015 | 11.024 | 11.007 |
| t(3) | Corr | 5.263 | 11.886 | 11.722 | 5.088 | 84.830 | 299.019 | 5.257 | 11.812 | 11.533 | 4.947 | 12.602 | 12.232 |
| | Ind | 5.148 | 11.761 | 11.593 | 5.056 | 81.398 | 515.612 | 5.065 | 10.911 | 10.809 | 4.566 | 11.426 | 11.042 |
| chisq(2) | Corr | 5.491 | 12.141 | 11.698 | 4.836 | 7.866 | 7.915 | 5.439 | 11.931 | 13.138 | 5.102 | 11.828 | 12.915 |
| | Ind | 5.386 | 11.972 | 10.883 | 4.791 | 7.452 | 7.496 | 4.970 | 11.352 | 12.621 | 4.759 | 10.951 | 12.243 |
| Only missing data is considered | | | | | | | | | | | | | |
| N(0,1) | Corr | 6.094 | 12.749 | 12.281 | 5.710 | 8.603 | 9.815 | 6.345 | 12.928 | 13.364 | 5.795 | 13.759 | 14.586 |
| | Ind | 5.797 | 12.207 | 11.336 | 5.591 | 8.234 | 9.060 | 6.042 | 11.925 | 12.453 | 5.478 | 13.472 | 13.772 |
| t(3) | Corr | 6.271 | 14.186 | 12.715 | 6.367 | 90.723 | 325.668 | 6.149 | 14.343 | 13.162 | 5.606 | 13.309 | 14.476 |
| | Ind | 6.227 | 13.442 | 11.940 | 5.996 | 83.167 | 596.296 | 5.745 | 13.656 | 12.091 | 5.319 | 13.288 | 14.329 |
| chisq(2) | Corr | 6.518 | 13.483 | 12.027 | 5.740 | 8.412 | 8.176 | 6.645 | 14.498 | 15.077 | 6.493 | 12.262 | 14.992 |
| | Ind | 6.044 | 12.895 | 11.039 | 5.639 | 7.978 | 7.678 | 6.627 | 13.764 | 14.601 | 6.468 | 11.683 | 14.744 |
| Measurement errors and missing data are considered | | | | | | | | | | | | | |
| N(0,1) | Corr | 8.961 | 18.213 | 18.330 | 8.622 | 12.715 | 14.217 | 8.990 | 19.564 | 19.556 | 9.063 | 19.433 | 20.847 |
| | Ind | 8.894 | 18.078 | 18.309 | 8.426 | 12.589 | 14.151 | 8.911 | 19.225 | 19.303 | 8.887 | 19.091 | 20.808 |
| t(3) | Corr | 8.783 | 19.912 | 18.778 | 9.563 | 132.242 | 469.511 | 9.615 | 20.398 | 18.804 | 8.027 | 20.251 | 20.378 |
| | Ind | 8.503 | 19.636 | 18.567 | 9.494 | 132.018 | 669.399 | 9.565 | 20.318 | 18.468 | 8.000 | 20.160 | 20.357 |
| chisq(2) | Corr | 9.600 | 19.553 | 18.533 | 8.360 | 12.316 | 12.689 | 9.781 | 20.766 | 22.066 | 9.068 | 18.703 | 21.581 |
| | Ind | 9.434 | 19.548 | 18.197 | 8.286 | 12.281 | 12.574 | 9.610 | 20.727 | 21.693 | 8.943 | 18.591 | 21.558 |

Similarly, the Atan penalty performs better than other penalty methods in the penalty term and tends to lead to simpler modelselection.

Table 3: The systematic names of selected gene for h=(0.1,1,10)

| textbfmethod | **systematic names** | | | | |
|---|---|---|---|---|---|
| scad(h=0.1) | Contig15674_RC | NM_002614 | NM_001730 | Contig52319_RC | NM_014204 |
| scad(h=1) | Contig19877_RC | NM_020182 | Contig62909_RC | Contig15674_RC | NM_006157 |
| scad(h=10) | NM_002614 | Contig19877_RC | Contig15674_RC | AJ009936 | NM_003359 |
| mcp(h=0.1) | Contig15674_RC | NM_014312 | NM_001730 | NM_004469 | NM_001807 |
| mcp(h=1) | Contig15674_RC | Contig19877_RC | NM_002614 | Contig12076_RC | AB033080 |
| mcp(h=10) | NM_001730 | Contig15674_RC | Contig19877_RC | AB011168 | NM_006409 |
| lasso(h=0.1) | Contig15674_RC | NM_020190 | NM_002614 | M34671 | Contig48934_RC |
| lasso(h=1) | NM_002614 | Contig19877_RC | Contig12076_RC | NM_001730 | Contig15674_RC |
| lasso(h=10) | Contig15674_RC | NM_020190 | NM_002614 | AB011168 | M34671 |
| atan(h=0.1) | Contig15674_RC | NM_002614 | Contig12076_RC | Contig19877_RC | Contig20889_RC |
| atan(h=1) | Contig29156_RC | NM_002614 | Contig15674_RC | Contig44793 | Contig52319_RC |
| atan(h=10) | Contig19877_RC | Contig15674_RC | NM_002614 | Contig62909_RC | NM_012454 |

## 6 Conclusions

For high-dimensional data with data missing and measurement error, we propose a robust variableselection method. In previous studies, when dealing with this kind of problem, the classical statistical methods often cause large deviation. We use the exponential square loss function to make the model robust, and use the inverse probability weighting method and orthogonal regression to solve this problem. In our experiment, covariates do not need to meet specific distribution rules. We also studied the theoretical nature of this problem. In addition, we conducted Monte Carlo simulation experiments to study the effect of the model under different conditions, respectively discussed the error under the condition of only considering measurement error, only considering data missing, and neither, and compared the effect of some traditional penalty functions. The experimental results show that our method has better effect than traditional methods in processing high-dimensional data with measurement errors and missing data.

Table 4: Analysis of cancer data for h=(0.1,1,10)

| **measure** | **scad** | | **map** | | **lasso** | | **atan** | |
|---|---|---|---|---|---|---|---|---|
| | **bias** | **size** | **bias** | **size** | **bias** | **size** | **bias** | **size** |
| Taking into account both measurement errors and missing data | | | | | | | | |
| h=0.1 | 2.77 | 101.77 | 2.69 | 91.44 | 2.77 | 98.56 | 2.44 | 80.33 |
| h=1 | 2.89 | 98.67 | 2.74 | 94.32 | 2.89 | 102.33 | 2.75 | 81.34 |
| h=10 | 3.01 | 97.56 | 2.91 | 93.70 | 3.01 | 96.88 | 2.80 | 79.24 |
| Only measurement errors is considered | | | | | | | | |
| h=0.1 | 2.92 | 105.15 | 2.71 | 94.38 | 3.02 | 106.02 | 2.82 | 80.43 |
| h=1 | 3.21 | 99.15 | 2.79 | 101.00 | 2.93 | 108.37 | 3.14 | 86.60 |
| h=10 | 3.46 | 104.86 | 3.18 | 100.77 | 3.07 | 104.69 | 3.21 | 82.12 |
| Only consider data missing | | | | | | | | |
| h=0.1 | 3.06 | 106.93 | 2.82 | 95.98 | 3.10 | 107.49 | 2.83 | 80.54 |
| h=1 | 3.36 | 99.30 | 2.85 | 101.18 | 3.05 | 109.96 | 3.29 | 88.49 |
| h=10 | 3.48 | 106.22 | 3.20 | 101.04 | 3.18 | 106.14 | 3.31 | 82.34 |
| Measurement errors and data missing are not considered | | | | | | | | |
| h=0.1 | 4.57 | 161.81 | 4.63 | 145.54 | 4.84 | 165.27 | 4.73 | 125.21 |
| h=1 | 5.49 | 154.76 | 4.67 | 154.39 | 5.04 | 165.29 | 5.00 | 131.97 |
| h=10 | 5.76 | 164.10 | 5.51 | 152.90 | 5.33 | 161.04 | 4.96 | 127.48 |